\documentclass[%
reprint,
superscriptaddress,
amsmath,amssymb,
aps,
]{revtex4-2}

\usepackage{dsfont}

\usepackage{calc}
\usepackage{accents}
\newcommand{\vardbtilde}[1]{\tilde{\raisebox{0pt}[0.85\height]{$\tilde{#1}$}}}
\usepackage{graphicx}
\usepackage{dcolumn}
\usepackage{bm}
\usepackage{tikz}
\usetikzlibrary{quantikz2}
\usepackage{nicematrix}
\usepackage{array}
\usepackage{makecell}
\usepackage{caption}
\usepackage{subcaption}
\usepackage{algorithm2e}
\RestyleAlgo{ruled}
\SetKwComment{Comment}{/* }{ */}
\SetKwInOut{Input}{Input}
\SetKwInOut{Output}{Output}
\SetKwFunction{FMain}{Main}
\SetKwProg{Fn}{Function}{:}{}
\SetKwInOut{Parameter}{Parameter}
\usepackage[pdftex,
            pdftitle={Greedy vs VQA},
            pdfauthor={Tim Schwägerl, Yahui Chai, Tobias Hartung, Karl Jansen, Stefan Kühn},
            bookmarks,
            colorlinks,
            linkcolor=myblue,
            citecolor=mymagenta,
            menucolor=black,
            urlcolor=myblue,
            plainpages=false,
            pdfpagelabels,
            hypertexnames=false]{hyperref}
\definecolor{mymagenta}{RGB}{200, 0, 100}
\definecolor{myblue}{RGB}{45, 48, 146}
\definecolor{mypurple}{RGB}{200, 112, 255}

\begin{document}

\preprint{APS/123-QED}

\title{Benchmarking Variational Quantum Algorithms for Combinatorial Optimization in Practice}

\author{Tim Schw\"agerl}
\affiliation{Institut f\"ur Physik, Humboldt-Universit\"at zu Berlin, Newtonstr. 15, 12489 Berlin, Germany}
\affiliation{CQTA, Deutsches Elektronen-Synchrotron DESY, Platanenallee 6, 15738 Zeuthen, Germany}
\author{Yahui Chai}
\affiliation{CQTA, Deutsches Elektronen-Synchrotron DESY, Platanenallee 6, 15738 Zeuthen, Germany}
\author{Tobias Hartung}
\affiliation{Northeastern University - London, Devon House, St Katharine Docks, London, E1W 1LP, United Kingdom}
\affiliation{Khoury College of Computer Sciences, Northeastern University, \#202, West Village Residence Complex H, 440 Huntington Ave, Boston, MA 02115, USA}
\author{Karl Jansen}
\affiliation{CQTA, Deutsches Elektronen-Synchrotron DESY, Platanenallee 6, 15738 Zeuthen, Germany}
\affiliation{Computation-Based Science and Technology Research Center, The Cyprus Institute, 20 Kavafi Street, 2121 Nicosia, Cyprus}
\author{Stefan K\"uhn}
\affiliation{CQTA, Deutsches Elektronen-Synchrotron DESY, Platanenallee 6, 15738 Zeuthen, Germany}

\date{\today}

\begin{abstract}
    Variational quantum algorithms and, in particular, variants of the varational quantum eigensolver have been proposed to address combinatorial optimization (CO) problems. Using only shallow ansatz circuits, these approaches are deemed suitable for current noisy intermediate-scale quantum hardware. However, the resources required for training shallow variational quantum circuits often scale superpolynomially in problem size. In this study we numerically investigate what this scaling result means in practice for solving CO problems using Max-Cut as a benchmark. For fixed resources, we compare the average performance of training a shallow variational quantum circuit, sampling with replacement, and a greedy algorithm starting from the same initial point as the quantum algorithm. We identify a minimum problem size for which the quantum algorithm can consistently outperform sampling and, for each problem size, characterize the separation between the quantum algorithm and the greedy algorithm. Furthermore, we extend the average case analysis by investigating the correlation between the performance of the algorithms by instance. Our results provide a step towards meaningful benchmarks of variational quantum algorithms for CO problems for a realistic set of resources. 
\end{abstract}

\maketitle

\section{\label{sec:intro}Introduction}
The goal of combinatorial optimization (CO) is finding an optimal or close to optimal solution from a finite set of solution candidates. 
CO problems have many applications of practical relevance for industry, such as supply chain optimization~\cite{Eskandarpour2015}, logistics~\cite{Sbihi2012}, and designing computer chips~\cite{Barahona1988}. 
Moreover, many problems in physics can be formulated as CO problems, for example, the reconstruction of particle trajectories in collider experiments~\cite{Saito2020, Gray2021, Zlokapa2021, Funcke2023, Schwaegerl2023, Crippa2023, Okawa2023}. 
Common CO problems are NP-hard~\cite{Karp1972}, which means that it is not expected to solve them efficiently, neither on a classical nor on a quantum computer. 
In practice, CO problems are routinely tackled by classical heuristic algorithms that provide approximate solutions. Since even small improvements in terms of accuracy or run-time would have a large impact on business and science, a significant effort is put into designing better algorithms for CO.

Quantum computing offers algorithms that provide theoretical speedups over the best known classical algorithms for very specific problems, such as factoring~\cite{Shor1994} and unstructured search~\cite{Grover1996}. 
It is an open question if they provide speedups for CO, which motivates many theoretical and empirical studies. In the era of noisy intermediate-scale quantum (NISQ) computers~\cite{Preskill2018}, most approaches to CO problems belong to the class of variational quantum algorithms (VQAs). VQAs are hybrid quantum-classical algorithms that use computations on classical computers to train the variational parameters of parameterized ansatz circuits, respecting the limitations of NISQ hardware. Prominent examples are the quantum approximate optimization algorithm (QAOA)~\cite{Farhi2014}, the variational quantum eigensolver (VQE)~\cite{Peruzzo2014} and numerous variations of both algorithms, which are reviewed in Ref.~\cite{Blekos2024}. In addition to algorithmic developments, there exist many proof of principle demonstrations and benchmarks of VQAs for CO problems~\cite{fga_vqe, Stollenwerk2021, Chai2023, Chai2023a, Hancock2023, amaro_case_2022, Nannicini2019}.

VQAs are often not trainable, due to cost concentration~\cite{Arrasmith2022} and poor local minima~\cite{Anschuetz2022}.
This means that the number of samples required to train the quantum circuit scales superpolynomially in problem size.  
In this study we numerically investigate what this scaling result means in practice for solving small instances of the prominent Max-Cut problem on 3-regular graphs. We compare the average performance of three algorithms that are all based on repeatedly evaluating the Max-Cut objective function, which allows for a direct comparison:
training a variational quantum circuit, sampling with replacement, and a greedy algorithm starting from the same initial point as the quantum algorithm. 
Our results aid the understanding of average case performance metrics for VQAs, such as the approximation ratio and the success probability. Furthermore, we check if good initial points for the greedy algorithm are also good initial points for the VQA by analyzing the correlation between the performance of both algorithms by instance.

This paper is organized as follows. In Sec.~\ref{sec:max-cut} we describe how we construct the instances of the Max-Cut problem that we use for our benchmark. Performance metrics for comparing the different algorithms are defined in Sec.~\ref{sec:metrics}. In particular, we introduce a metric to gain intuition for the meaning of the average difference in approximation ratio and an approach using a binned statistic to investigate the correlation of the performance of the algorithms by instance.
In Sec.~\ref{sec:algorithms} we describe the specific VQA of our study, the greedy algorithm, and the baseline sampling procedure. In particular, we define what we mean by starting the VQA and the greedy algorithm from the same initial point. We provide a short overview of related work in Sec.~\ref{sec:related}.
The setup of our numerical experiments and the results are presented in Sec.~\ref{sec:experiments}.
We conclude with a discussion and an idea to extend our benchmarks to quantum problems in Sec.~\ref{sec:dicussion}. 
\section{\label{sec:co}Combinatorial optimization}
In the following we briefly introduce the Max-Cut problem that we use as a benchmark, and discuss the various performance metrics for characterizing the different algorithms.
\subsection{\label{sec:max-cut}Max-Cut}
The goal of the Max-Cut problem is to divide the nodes $V$ of an undirected weighted graph $G = (V, E)$ with edges $E$ into two sets such that the sum of the edge weights between the two sets is maximal. Formally, for a graph $G = (V, E)$ with $|V|=N+1$ nodes and edge weights $w_{ij}>0$ for $(i, j) \in E$ the problem is to maximize the objective function
\begin{align}
    \vardbtilde{O}(\boldsymbol{x}) = \sum_{(i, j) \in E} w_{ij} [ x_i(1-x_j) + x_j(1-x_i)].
\end{align} 
by assigning $x_i=0$ or $x_i=1$ to each node $i$. We follow the convention $j<i$ for labeling the edges $(i, j) \in E$. An assignment $\boldsymbol{x} = (x_1,\dots, x_{N+1})$ divides the graph into two sets of nodes according to their labels. Maximizing the objective function corresponds to maximizing the sum of the edge weights between the two sets. The problem has a symmetry under interchanging the labels $0 \to 1$ and $1 \to 0$. Following Ref.~\cite{Amaro2022}, we remove this symmetry by setting $x_1=0$, leading to the objective function
\begin{multline}
    \tilde{O}(\boldsymbol{x}) = \sum_{(i, j=1) \in E}w_{ij}x_i \\
    + \sum_{(i \neq 1, j \neq 1) \in E} w_{ij} [ x_i(1-x_j) + x_j(1-x_i)].
\end{multline}
The Max-Cut problem on a graph with $N+1$ nodes is now encoded as an objective function of $N$ binary variables. Finding the optimal solution $\boldsymbol{x}^*= \mathrm{arg max}\,\vardbtilde{O}(\boldsymbol{x})$ is NP-hard. Finding a solution $\boldsymbol{x}$ with approximation ratio $\alpha =\vardbtilde{O}(\boldsymbol{x})/\vardbtilde{O}(\boldsymbol{x}^*)>16/17\approx  0.9412$ is also NP-hard~\cite{Haastad2001}. This makes Max-Cut a suitable problem for studying algorithms for CO. The Goemans-Williamson (GW) algorithm, the best-known classical semidefinite programming (SDP) algorithm for Max-Cut, can find a solution with approximation ratio of $\alpha \geq 0.87856$ in polynomial time~\cite{Goemans1995}.

We follow a similar approach as Ref.~\cite{Amaro2022} to construct Max-Cut instances for our numerical experiments. 
For each $N \in \{11, 21, 31, 41, 51, 61\}$, we define 25 Max-Cut instances on random 3-regular simple undirected and connected graphs with weights $w_{ij}$ drawn uniformly from $(0, 1]$. We divide the objective function $\tilde{O}$ of every instance by the objective value $O_{\mathrm{GW}}$ achieved by a single run of the GW SDP algorithm using its implementation in Qiskit~\cite{qiskit2024}. This defines the objective function $O$ we use in our numerical experiments:
\begin{align}
    O(\boldsymbol{x}) = \frac{\tilde{O}(\boldsymbol{x})}{O_{GW}}.
\end{align}
Now, the objective functions are re-scaled to intervals $[0, \beta]$ where the lower bound 0 is given by the trivial cut $\boldsymbol{x}=0$.
The objective function $O$ is the same as the ratio of the corresponding approximation ratios. Since the GW algorithm gives an approximation ratio of 0.87856 in the worst case and an approximation ratio of 1 in the best case, it is evident that $1 \leq \beta \leq 1/0.87856$.
This procedure eliminates possible impacts of differing scales of the objective function on the performance of the algorithms we benchmark.  
\subsection{\label{sec:metrics} Performance metrics}
To quantify an algorithm's ability to find an approximate solution $\boldsymbol{x}_{\mathrm{max}}$ for the Max-Cut problem, we use the approximation ratio:
\begin{align}
    \alpha = \frac{O(\boldsymbol{x}_{\mathrm{max}})}{O(\boldsymbol{x}^*)}.
\end{align}
Throughout each algorithm that we run, we keep track of the assignment that produced the highest approximation ratio. The final approximation ratio is then defined as the highest approximation ratio observed during the runtime of the algorithm.
In particular, for the case of VQE we do not use the expected value of the objective function to determine the approximation ratio, which would correspond to an average over multiple assignments, if the current ansatz state does not represent a computational basis state. Instead, we use the best assignment $\boldsymbol{x}_{\mathrm{max}}$ obtained by the measurements throughout the VQE process.
The approximations ratios achieved by the VQE in our numerical experiments follow a highly non-normal distribution over problem instances and initial points. This means that its mean value does not represent the approximation ratio of a typical VQE instance. Furthermore, the sample standard deviation does not include most VQE instances. Thus, we compute the standard error of the mean (SEM) instead of the sample standard deviation, rating how well the sample mean represents the population mean. To aid in understanding what a certain average difference in the approximation ratio between two algorithms means, we use an additional statistical method. We compute mean probabilities and, using Wilson's score method~\cite{Wilson1927} implemented in Python's statsmodels module~\cite{seabold2010statsmodels}, 95\% confidence intervals for an algorithm achieving a higher approximation ratio than another algorithm. This metric does not include information about the value of the difference but gives an intuitive understanding complementing the average difference in approximation ratio. In this metric, a value of 1/2 means that the algorithms perform equally well.

For small problems, the VQE is able to find the exact solution. To quantify its ability to do so, we count the number of successful runs where success is defined as observing the exact solution at least once when running the algorithm. Then, we compute the mean success probability and, again using Wilson's score method, the 95\% confidence interval for success. When benchmarking the VQE, this metric is only useful for small problems because it decays rapidly with problem size. For larger problems, one cannot expect to find the exact solution and has to rely on the approximation ratio as a performance metric. 

To obtain an understanding of the algorithms beyond the typical average case studies, we analyze the correlation of the approximation ratios achieved by different algorithms by instance. We do so using binned statistics on the instances as outlined in the following. An instance is defined by the problem instance and, if the algorithm accepts an initial point, by the initial point. First, we compute the approximation ratios achieved by the algorithms for every instance. Then, we group the instances into small intervals of equal size in approximation ratio achieved by one algorithm. The $x$-value is defined by the sample mean of these approximation ratios and its sample standard deviation, which is small by construction. Now, the $y$-value is given by the sample mean and standard deviation of the approximation ratio achieved by another algorithm for the same instances. This method allows for investigating if instances that are hard for one algorithm are also hard for another algorithm.

\section{\label{sec:algorithms}Algorithms}
In this section we introduce the three algorithms we compare in our study: training a shallow parametric quantum circuit using VQE, sampling with replacement, and a greedy algorithm starting from the same initial point as the quantum algorithm.
\subsection{\label{sec:vqe}Variational quantum eigensolver}
The VQE is a hybrid quantum-classical algorithm that was originally proposed for computing ground states of molecular Hamiltonians~\cite{Peruzzo2014}. The expectation value of the Hamiltonian is evaluated on a quantum device using a parametrized circuit as a variational ansatz. The parameters are then updated utilizing a classical optimization algorithm such that the Hamiltonian's expectation value decreases. Running this quantum-classical feedback loop iteratively until certain convergence criteria are matched or when a maximum number of $N_{\mathrm{iter}}$ iterations is reached, one obtains an approximation for the ground state of the Hamiltonian. In our study we use the VQE to maximize the value of an objective function which is equivalent to minimizing the corresponding Hamiltonian.

In the context of CO, an alternative view on the algorithm is preferable because the solution candidates are computational basis states.  
The quantum circuit $M$, which we assume to include projective measurements in the computational basis at the end, is a model that maps sets of $N_\mathrm{params}$ parameters $\boldsymbol{\vartheta}$ to computational basis states of $N$ qubits. 
Here we consider without loss of generality parametric gates of the form $\exp(i\vartheta_k \mathcal{P})$, where $\mathcal{P}$ is a Pauli string on $N$ qubits, $\mathcal{P}\in\{\mathds{1},X,Y,Z\}^{\otimes N}$. Hence, the parameters can be restricted to $[0,2\pi)$, and the quantum circuit corresponds to
\begin{align}
    M:[0, 2\pi)^{N_{\mathrm{params}}} \to \{0,1\}^N.
\end{align}
The cost function $C$, the target of the the classical optimization algorithm, computes a single cost value for $N_\mathrm{shots}$ computational basis states
\begin{align}
    C: (\{0,1\}^N)^{N_{\mathrm{shots}}} \to \mathbb{R}.
\end{align}
The classical optimization algorithm $OA$ uses this value and the corresponding set of parameters to calculate a new set of parameters~\footnote{Note that for simplicity of notation, we assume the optimization algorithm only takes the parameter values and the current cost function value as in input. Additional arguments, which would be required, e.g., for gradient-based optimization algorithms would not affect any of the arguments presented in the following.}
\begin{align}
    OA: [0, 2\pi)^{N_{\mathrm{params}}} \times \mathbb{R} \to [0, 2\pi)^{N_{\mathrm{params}}}.
\end{align}
The VQE algorithm in terms of these definitions is described in Algorithm~\ref{alg:vqe}.
\begin{algorithm}
\caption{Variational Quantum Eigensolver}\label{alg:vqe}
\Input{Quantum circuit $M$, initial parameter vector $\boldsymbol{\vartheta}$, objective function $O$, cost function $C$, optimization algorithm $OA$}
\Parameter{Maximum number of iterations $N_{\mathrm{iter}}$, maximum number of shots $N_{\mathrm{shots}}$}
\Output{Binary vector $\boldsymbol{x}_{\mathrm{max}} \in \{0,1\}^N$}
$O_{\mathrm{max}} \gets -\infty$ \;
\For{$\mathrm{iteration} \gets 1$ \KwTo $N_{\mathrm{iter}}$}{
    $X \gets \text{empty list}$\;
    \For{$\mathrm{shot} \gets 1$ \KwTo $N_{\mathrm{shots}}$}{
        $\boldsymbol{x}_{\mathrm{update}} \gets M(\boldsymbol{\vartheta})$\;
        $X\mathrm{.append}(\boldsymbol{x}_{\mathrm{update}})$\;
        $O_{\mathrm{update}} \gets O(\boldsymbol{x}_{\mathrm{update}})$\;
        \If{$O_{\mathrm{update}} > O_{\mathrm{max}}$}{
            $\boldsymbol{x}_{\mathrm{max}} \gets \boldsymbol{x}_{\mathrm{update}}$\;
            $O_{\mathrm{max}} \gets O_{\mathrm{update}}$\;
        }
        }
    $\boldsymbol{\vartheta} \gets OA(\boldsymbol{\vartheta}, C(X))$\;
    }
\end{algorithm}
\subsection{\label{sec:vqe_details}Hyper-parameter choices for the VQE}
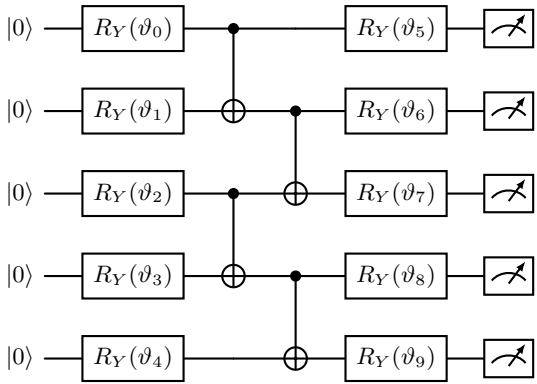
\begin{figure}
    \begin{tikzpicture}
    \node[scale=1]{
        \begin{quantikz}
            \lstick{\ket{0}}&\gate{R_Y(\vartheta_0)}&\ctrl{1}&&\gate{R_Y(\vartheta_5)}&\meter{}\\
            \lstick{\ket{0}}&\gate{R_Y(\vartheta_1)}&\targ{}&\ctrl{1}&\gate{R_Y(\vartheta_6)}&\meter{}\\
            \lstick{\ket{0}}&\gate{R_Y(\vartheta_2)}&\ctrl{1}&\targ{}&\gate{R_Y(\vartheta_7)}&\meter{}\\
            \lstick{\ket{0}}&\gate{R_Y(\vartheta_3)}&\targ{}&\ctrl{1}&\gate{R_Y(\vartheta_8)}&\meter{}\\
            \lstick{\ket{0}}&\gate{R_Y(\vartheta_4)}&&\targ{}&\gate{R_Y(\vartheta_9)}&\meter{}
        \end{quantikz}
    };
    \end{tikzpicture}
\caption{The circuit structure of the VQE for 5 variables. While all statements of our study refer to a single layer of this type, more complex algorithms use it as a building block.}
\label{fig:circuit}
\end{figure}

In our study we investigate the performance of a simple hardware-efficient circuit architecture that is suitable for noisy intermediate-scale quantum hardware. The circuit starts with a register of $N$ qubits encoding $N$ binary variables initialized in the state $|0\rangle ^{\otimes N}$. Then, a single layer of parameterized $R_Y$ rotation gates acts on the qubits. The exclusive use of $R_Y$ rotations results in a low gate count and a circuit that generates real amplitudes. The initial layer is sufficient to express every computational basis state but the corresponding cost landscape suffers from numerous local minima. To generate a possibly more favorable cost landscape, entangling CNOT gates are added in a brick like pattern as shown in Figure~\ref{fig:circuit}. This pattern enables the execution of multiple CNOT operations in parallel. The circuit is completed with another layer of $R_Y$ rotations and a measurement of all qubits in the computational basis. 
Since the VQE was proposed to estimate ground-state energies, its original cost function is the energy expectation value. In the context of CO, this would lead to the cost function $C$ which is the full sample mean of the objective function for $N_{\mathrm{shots}}$ computational basis states $\boldsymbol{x}_k$ that we denote with $X$:
\begin{align}
    C(X) = \frac{1}{N_{\mathrm{shots}}} \sum_{k=1}^{N_{\mathrm{shots}}} O(\boldsymbol{x}_k)
\end{align}
In our study, we use the conditional value at risk (CVaR) as a cost function, $C_{\mathrm{CVaR}}$, that was proposed to enhance the performance of VQAs for CO~\cite{Barkoutsos2020}. Assuming that the objective values $O(\boldsymbol{x}_k)$ are sorted in non-increasing order, the CVaR cost function only considers the fraction $\gamma$ of large objective values:
\begin{align}
    C_{\mathrm{CVaR}}(X) = \frac{1}{\lceil\gamma N_{\mathrm{shots}}\rceil} \sum_{k=1}^{\lceil\gamma N_{\mathrm{shots}}\rceil} O(\boldsymbol{x}_k).
    \label{eq:cvar}
\end{align}
The motivation for the CVaR cost function lies in the observation that, for CO, quantum states with large components with high objective values are favorable over states with possibly larger mean objective values but smaller components with high objective values. The case $\lceil\gamma N_{\mathrm{shots}}\rceil = 1$ results in an optimization with respect to the maximal observed objective value, while the case $\gamma=1$ retrieves the sample mean, which corresponds to the original VQE approach. The former leads to a discontinuous optimization landscape, making the choice of $\gamma$ an important hyper-parameter.
Throughout our numerical experiments we set $\gamma=0.1$ which we found to be close to optimal after testing multiple values. 

To update the parameters, we choose the gradient-free Constrained Optimization BY Linear Approximation (COBYLA) algorithm~\cite{Powell1998}. We use the default hyper-parameters in its SciPy implementation\cite{2020SciPy-NMeth}. 

Further important hyper-parameters are the maximal number of iterations $N_{\mathrm{iter}}$ and the number of measurements taken within one iteration $N_{\mathrm{shots}}$. We investigate the algorithms in the regime below $N_{\mathrm{evals}}=N_{\mathrm{iter}}N_{\mathrm{shots}}=10^6$ total objective function evaluations. By testing $N_{\mathrm{shots}}=10,10^2,10^3,10^4,10^5$ and the corresponding value of $N_{\mathrm{iter}}$, we found $N_{\mathrm{shots}}=10^3$ and $N_{\mathrm{iter}}=10^3$ to perform the best on average. This is the setup that we use throughout our study. 

We are aware that one does not expect the shallow VQE of our study to show competitive performance. In particular, we expect the resources required to sufficiently train the circuit of Fig.~\ref{fig:circuit} to scale superpolynomially with problem size because the shallow circuit leads to poor local minima. However, we aim at gaining intuition for what these scaling results mean in practice for fixed problem sizes. Furthermore, this type of circuit is used in more complex algorithms, such as the Filter-VQE algorithm~\cite{Amaro2022}, and as a building block of the Layer-VQE algorithm~\cite{Liu2022}. Our benchmarks can be applied straightforwardly to these algorithms and, in general, to all algorithms that rely on repeatedly evaluating the objective function. 
\subsection{\label{sec:sampling}Sampling with replacement}
In general, the VQE requires many iterations $N_{\mathrm{iter}}$ and measurements $N_{\mathrm{shots}}$ to converge on average. Through the connection $N_{\mathrm{evals}}=N_{\mathrm{iter}}N_{\mathrm{shots}}$, the VQE can be straightforwardly compared to other algorithms that rely on recurring evaluations of the objective function. This benchmark has a strong bias in favor of the quantum algorithm, because it relies on the assumption that generating a random computational basis state on a classical computer is as complex as executing the corresponding quantum circuit. The weakest competitor is uniformly sampling computational basis states with replacement, which does not make use of any structure of the problem. We choose sampling with replacement over sampling without replacement because the VQE does not directly use information about which computational basis states have been sampled before. The probability of sampling the optimal solution of a problem of size $N$ is given by $N_{\mathrm{evals}}/2^N$. In our study, we numerically assess the approximation ratio achieved by sampling with replacement. This simple benchmark can exclude the possibility that the VQE is just guessing solutions. 
\subsection{\label{sec:greedy}Greedy algorithm}
An alternative to training a probabilistic model to generate computational basis states is to define a fixed set of rules to generate states. A possible choice is a greedy algorithm that updates states in a way that locally maximizes the objective function. The input to the algorithm is a computational basis state $\boldsymbol{x}$. Then, the algorithm computes the objective function for all states that differ from the initial state $\boldsymbol{x}$ by exactly one bit flip, and accepts the state with the largest improvement in objective value as the initial state for the next iteration. If there is no further improvement or if the maximal number of evaluations of the objective function $N_{\mathrm{evals}}$ is reached, the algorithm is terminated. The procedure is explained in detail in Algorithm~\ref{alg:greedy}. We note that the evaluation of the objective function can be implemented  very efficiently for the greedy algorithm, because only the change of the objective function with respect to a single variable has to be computed. This means that the benchmark again has a bias in favor of the quantum algorithm.
\begin{algorithm}
\caption{Greedy algorithm}\label{alg:greedy}
\Input{Quantum circuit $M$, initial parameter vector $\boldsymbol{\vartheta}$, objective function $O$}
\Parameter{Maximum number of steps $N_{\mathrm{evals}}$}
\Output{Binary vector $\boldsymbol{x}_{\mathrm{max}} \in \{0,1\}^N$}
$\boldsymbol{x}_{\mathrm{max}} \gets M(\boldsymbol{\vartheta})$\;
$O_{\mathrm{max}} \gets O(\boldsymbol{x}_{\mathrm{max}})$\;
$\boldsymbol{x}_{\mathrm{temp}} \gets \boldsymbol{x}_{\mathrm{max}}$\;
    \While{$n_{\mathrm{evals}} \gets 1 < N_{\mathrm{evals}}$}{
            \DontPrintSemicolon $\boldsymbol{x}_{\mathrm{update}} \gets \text{ the single bit flip of } \boldsymbol{x}_{\mathrm{temp}} \text{ with}$\;
            \PrintSemicolon \Indp the largest objective function value\;
            \Indm $n_{\mathrm{evals}} \gets n_{\mathrm{evals}} + \text{number of single bit flips}$\;
            $O_{\mathrm{update}} \gets O(\boldsymbol{x}_{\mathrm{update}})$\;
            \eIf{$O_{\mathrm{update}} > O_{\mathrm{max}}$}{
                $\boldsymbol{x}_{\mathrm{max}} \gets \boldsymbol{x}_{\mathrm{update}}$\;
                $O_{\mathrm{max}} \gets O_{\mathrm{update}}$\;
                }{
                $\boldsymbol{x}_{\mathrm{temp}} \gets M(\boldsymbol{\vartheta})$\;
                $n_{\mathrm{evals}} \gets n_{\mathrm{evals}} + 1$\;
                $O_{\mathrm{temp}} \gets O(\boldsymbol{x}_{\mathrm{temp}})$\;
                \If{$O_{\mathrm{temp}} > O_{\mathrm{max}}$}{
                    $\boldsymbol{x}_{\mathrm{max}} \gets \boldsymbol{x}_{\mathrm{temp}}$\;
                    $O_{\mathrm{max}} \gets O_{\mathrm{temp}}$\;
                    }
                }
            }
\end{algorithm}

In the following we explain what we mean by starting the VQE and the greedy algorithm from the same initial point. For the VQE, the initial point is given by the quantum circuit and the initial parameters. The greedy algorithm, on the other hand, starts from a single computational basis state. We generate such a state by executing the quantum circuit for the same initial parameters as the VQE a single time. Then, we run the greedy update routine starting from this computational basis state. If no further improvement is possible, a new computational basis state is generated with the quantum circuit. Throughout this procedure, the algorithm keeps track of states it has already visited and breaks the loop when it reaches such a state. In that case a new initial state is generated using the quantum circuit. 

This method allows for comparing the performance of the VQE and the greedy algorithm on average for the same set of initial points. This is important because the performance of the VQE depends strongly on its initialization. Furthermore, by studying the correlation of the performance of the two algorithms by instance, we can understand if good initial points for the VQE are also good initial points for the greedy algorithm. 
\section{\label{sec:related}Related work}
In this section we summarize related work that we are aware of. Reference~\cite{Scriva2023} compares the performance of a QAOA and a VQE to sampling with replacement for ferromagnetic and disordered Ising chains. They probe the regime to up to around 20 variables where they do not observe a practical advantage of the quantum algorithms, in agreement with our numerical findings in Sec.~\ref{sec:experiments}. Furthermore, they propose a parameter initialization strategy to enhance the performance of the QAOA. 

In an extended performance analysis~\cite{MarinSanchez2024} of the Filter-VQE algorithm~\cite{Amaro2022} that also includes a comparison to sampling, the authors conclude that significant algorithmic developments are necessary to be competitive with state-of-the-art solvers, such as Gurobi~\cite{gurobi}. 

The authors of Ref.~\cite{Dupont2023} propose a quantum-enhanced greedy CO solver and compare it to its classical counterpart. The quantum version of the algorithm uses states obtained by the QAOA algorithm as starting points for the greedy procedure, while the classical version starts from a uniform distribution of all computational basis states. 

In Ref.~\cite{Sciorilli2024}, the authors propose a novel heuristics for quantum-inspired solvers that relies on encoding variables into correlations of Pauli-operators. Their encoding and the corresponding cost function lead to the best performance in experiment of a VQA for CO so far. They report average approximation ratios that are significantly better than a local search starting from randomly picked graph partitions. This is very similar to our comparison of the VQE to the greedy algorithm, which additionally uses the concept of starting both algorithms from the same initial point in the sense defined in Sec.~\ref{sec:greedy}.
\section{\label{sec:experiments}Numerical experiments}
We compare the performance of the three different algorithms described in Sec.~\ref{sec:algorithms} for the Max-Cut instances explained in Sec.~\ref{sec:max-cut}. For each of the six different problem sizes, we generate 25 instances and run the three different algorithms 10 times, for different initial points in case of the VQE and the greedy algorithm. This leads to a total of 4500 instances that we use to conduct average case convergence studies and to investigate the correlation of the performance of the algorithms. We use the Gurobi solver~\cite{gurobi} to obtain the optimal solutions of all Max-Cut instances. Note that the problem sizes considered in this study are trivial for commercial solvers, and Gurobi takes at most a few milliseconds to find the optimal solution and to prove that it is indeed optimal. 

For the VQE, we use the matrix product state simulation method of Qiskit Aer \cite{qiskit2024} to simulate it in an ideal setting without hardware noise. For convenience, we summarize the hyper-parameters of the VQE in Tab.~\ref{tab:vqe}, a detailed explanation of our choices was provided in Sec.~\ref{sec:vqe_details}.
\begin{table}[h]
    \centering
    \begin{tabular}{ | p{3cm} | p{5cm}| } 
        \hline
        Backend & Matrix product state simulator of Qiskit~\cite{qiskit2024} Aer without noise.\\ 
        \hline
        Circuit $M$  & $R_Y$ rotations and brick pattern of CNOT gates (Fig.~\ref{fig:circuit}).\\
        \hline
        Cost function $C$  & $C_{\mathrm{CVaR}}$ with $\gamma=0.1$ (Eq.~\ref{eq:cvar}).\\
        \hline
        Maximal number of iterations $N_{\mathrm{iter}}$  & 1000\\
        \hline
        Number of measurements per iteration $N_{\mathrm{shots}}$  & 1000\\
        \hline
        Optimization algorithm $OA$  & Constrained Optimization BY Linear Approximation (COBYLA) with the default hyper-parameters of its SciPy implementation~\cite{2020SciPy-NMeth}.\\
        \hline
    \end{tabular}
    \caption{Summary of the hyper-parameters of the VQE.}
    \label{tab:vqe}
\end{table}
The matrix product state method is well suited for the shallow circuit shown in Fig.~\ref{fig:circuit}, even for a large number of qubits.

For benchmarking sampling with replacement, we use the NumPy random number generator \cite{numpy} to generate computational basis states. We implemented the greedy algorithm with native Python data structures and NumPy. 
\subsection{\label{sec:average}Performance on average}
To compare the algorithms, we use the mean and the standard error of the mean (SEM) of the difference in approximation ratio, as defined in Sec.~\ref{sec:metrics}. Furthermore, we use the additional binned statistics introduced in the same section. For each problem size, we compute the differences for the 25 problem instances and, if the algorithm accepts an initial point, for the 10 initial points. For comparison to sampling, we use 10 independent runs that do evidently not depend on an initial point. Then, we compute the mean and SEM of these 250 differences. To study the convergence behavior of the algorithms, we apply this procedure after every iteration of the VQE. To compare the VQE with sampling and the greedy algorithm for the same number of generated computational basis states and calls of the objective function, we compare the $i\mathrm{th}$ iteration of the VQE to the best approximation ratio achieved after $i \times N_{\mathrm{shots}}$ steps of sampling or the greedy algorithm. 
\begin{figure}
    \begin{subfigure}{0.48\textwidth}
        \centering
        \includegraphics[width=\textwidth]{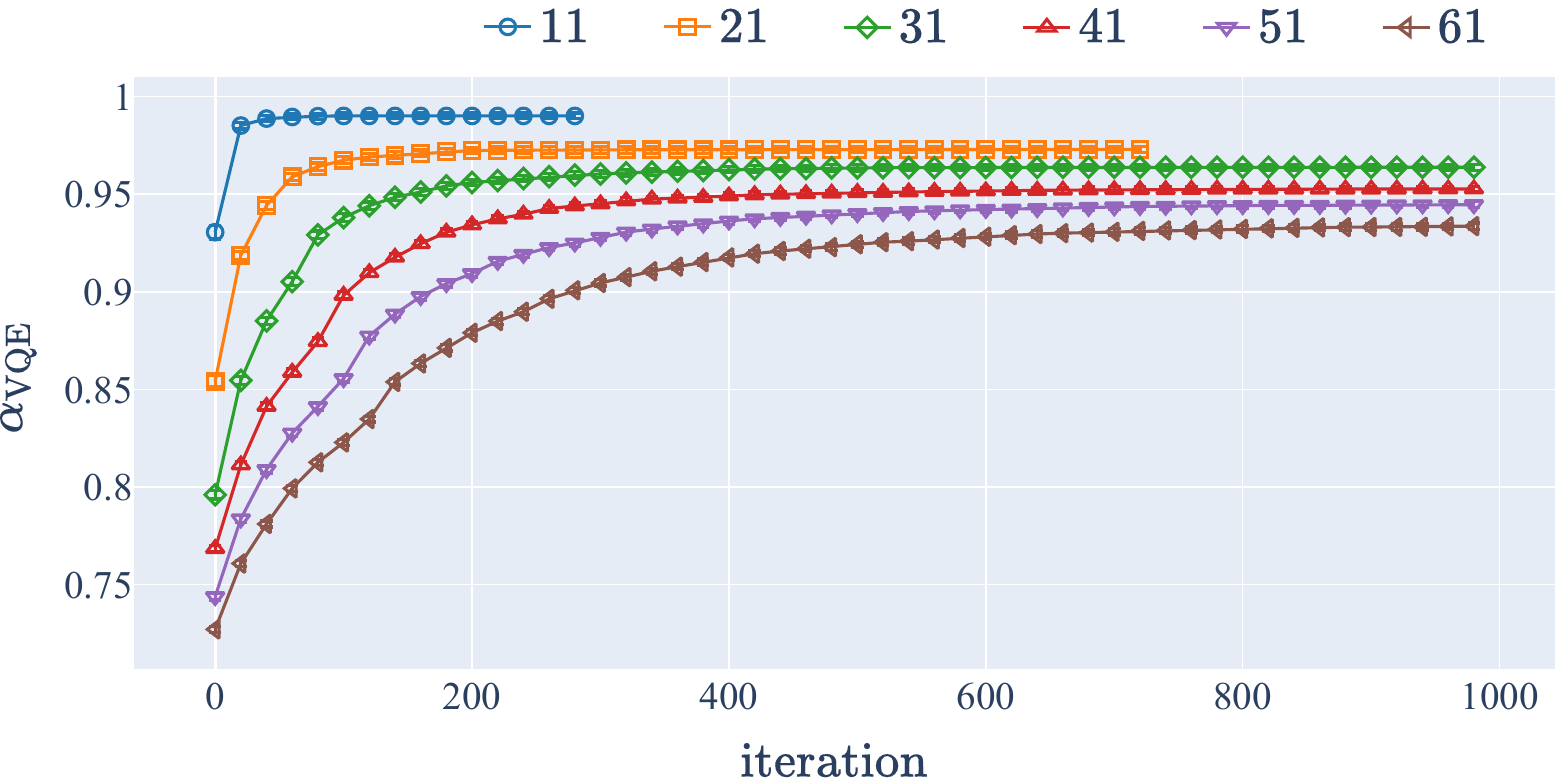}
        \caption{Mean and SEM of the approximation ratio achieved by the VQE for the Max-Cut problem on 3-regular graphs of different size.}
        \label{fig:approx_ratio_vqe}
    \end{subfigure}
    \hfill
    \begin{subfigure}{0.48\textwidth}
        \centering
        \includegraphics[width=\textwidth]{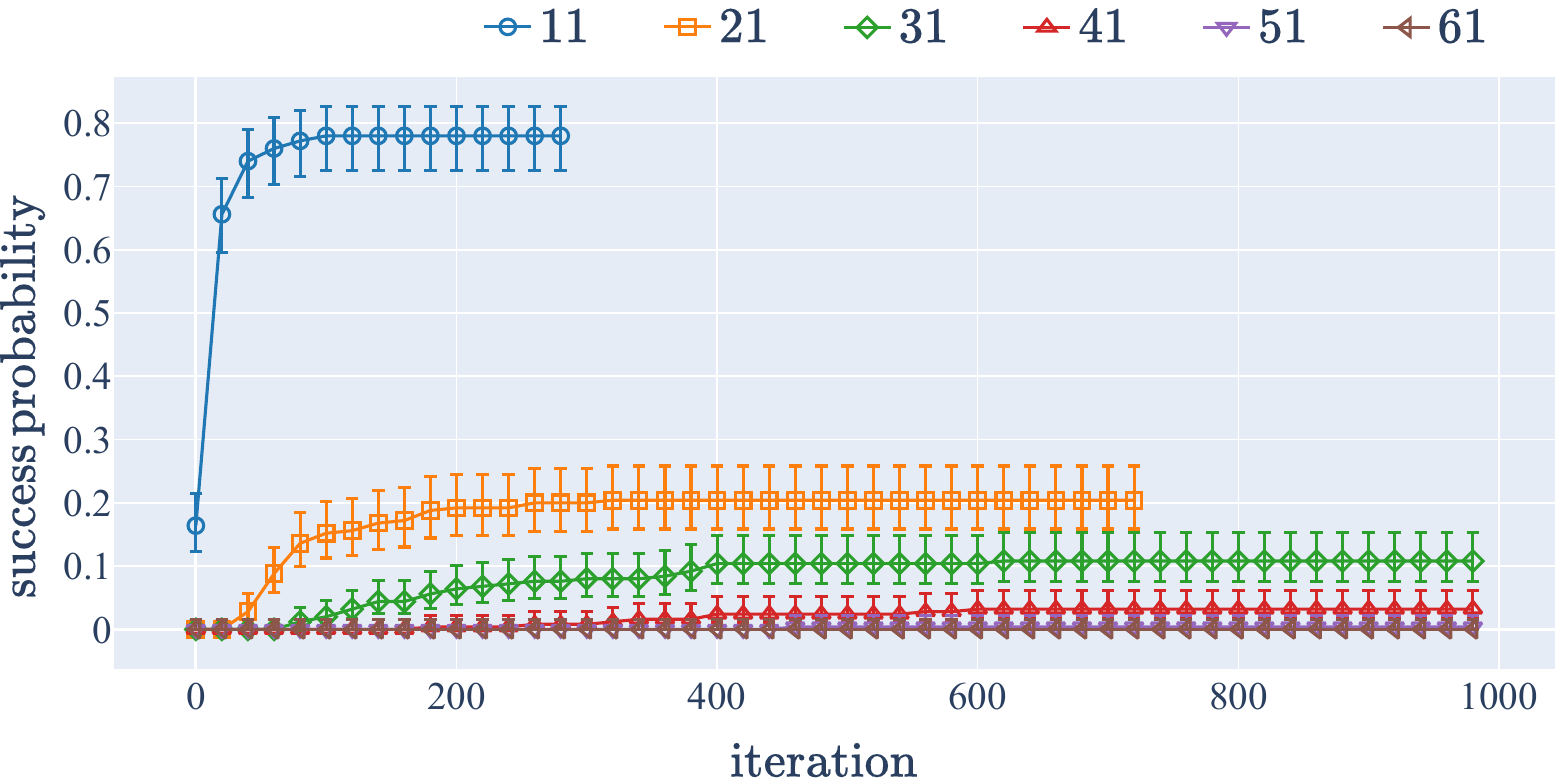}
        \caption{Mean probability and 95\% confidence interval of finding the exact solution. This metric is only useful for benchmarking this VQE for small problems because it decays rapidly with problem size.}
        \label{fig:success_prob_vqe}
    \end{subfigure}
    \hfill
    \caption{Frequently used performance metrics for VQAs applied to combinatorial optimization problems. 
    }
    \label{fig:vqe}
\end{figure}

\begin{figure}
    \begin{subfigure}{0.48\textwidth}
        \centering
        \includegraphics[width=\textwidth]{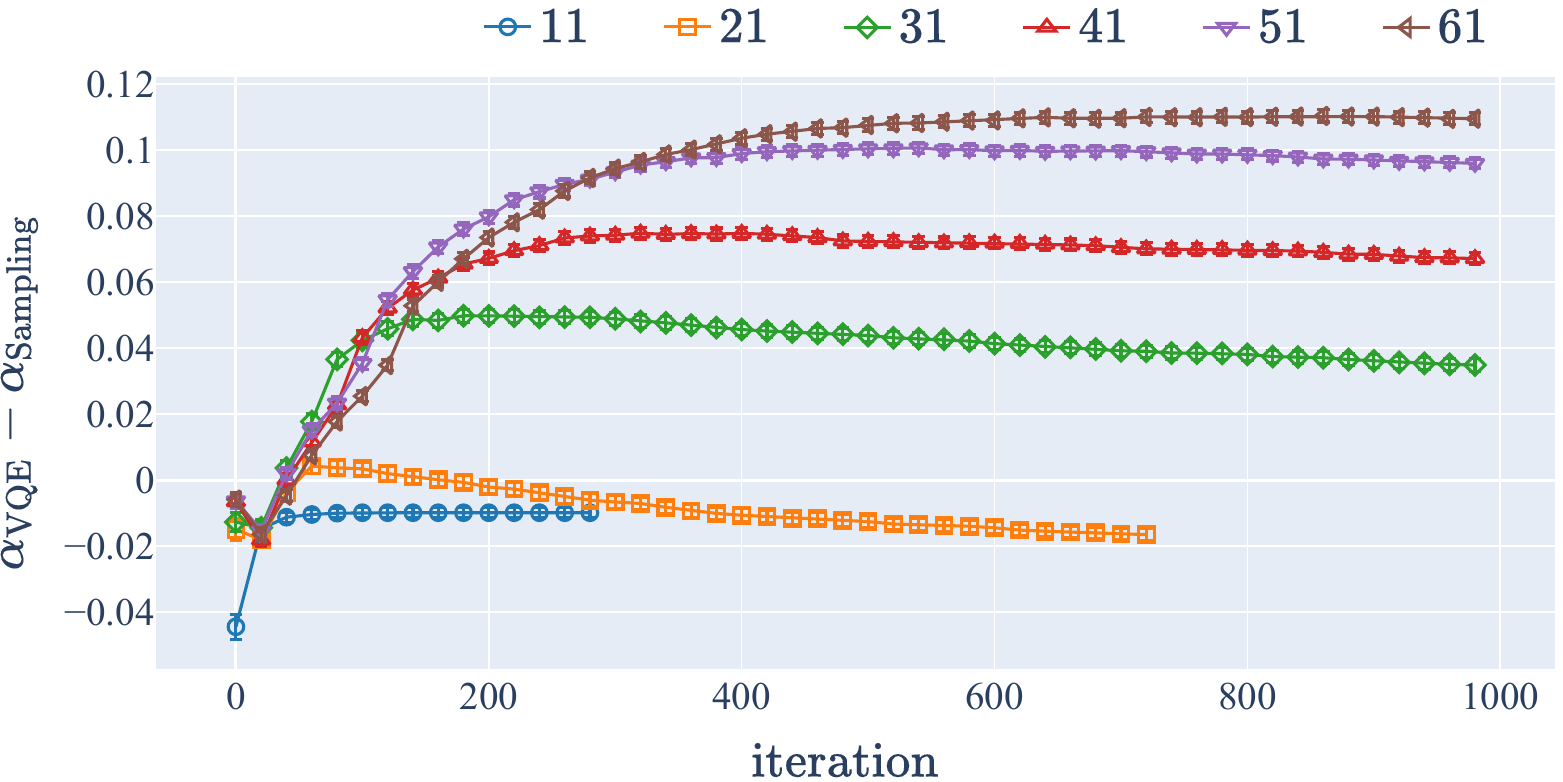}
        \caption{Mean and SEM of the difference in approximation ratio. For size 11 and a wide range of iterations for size 21, the VQE performs worse than sampling.}
        \label{fig:approx_ratio_sampling}
    \end{subfigure}
    \hfill
    \begin{subfigure}{0.48\textwidth}
        \centering
        \includegraphics[width=\textwidth]{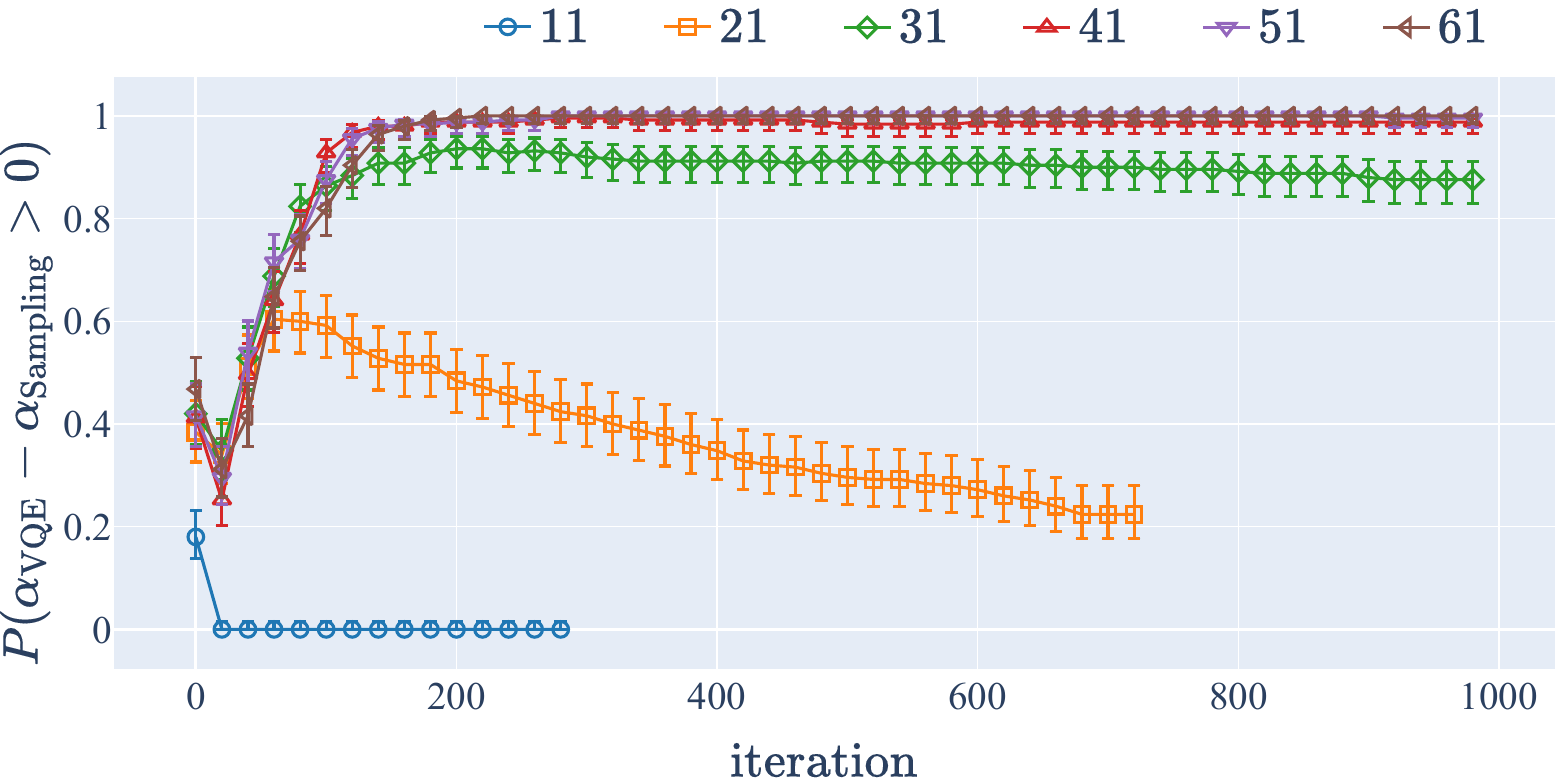}
        \caption{Mean probability and 95\% confidence interval of the VQE achieving a better approximation ratio than sampling. We propose this as an additional, intuitive metric for comparing algorithms.}
        \label{fig:better_approx_ratio_sampling}
    \end{subfigure}
    \hfill
    \caption{Average difference in performance metrics of the VQE and sampling with replacement as a function of the number of iteration for different problem sizes indicated by the different markers. As a guide for the eye, the different data points for each problem size are connected with a line. Even though the VQE achieves its best approximation ratio for size 11, there is only an advantage over sampling for larger problem sizes.}
    \label{fig:vqe_minus_sampling}
\end{figure}

First, we focus on the approximation ratio and the success probability achieved of the VQE which is shown in Fig.~\ref{fig:vqe}(a) and Fig.~\ref{fig:vqe}(b) respectively. In particular, for a problem size of 11, we observe an approximation ratio close to 1 and success probabilities well above 75\%.  Analyzing the approximation ratio of the VQE in comparison to the performance of sampling with replacement, we observe the picture in Fig.~\ref{fig:vqe_minus_sampling}(a). 
Now, it is evident that the high approximation ratio of the VQE for problem size 11 in Fig.~\ref{fig:vqe}(a) is misleading, because it is not better than randomly guessing solutions. A similar behavior is visible for problem size 21 where the VQE performs worse than sampling, except for a few early iterations. Our statistics in Fig.~\ref{fig:vqe_minus_sampling}(b) show that at around 50 iterations, the VQE achieves a higher approximation ratio than sampling for around 60\% of the instances. This statement is equivalent to sampling achieving an approximation ratio equal to or higher than the VQE for around 40\% of the instances and translates to an average difference in the approximation ratio of around 0.005 (see Fig.~\ref{fig:vqe_minus_sampling}(a)). At around 200 iterations, the two algorithms perform equally well on average, and for even larger numbers of iterations, sampling is consistently better than the VQE. 

A different picture emerges when one considers problem sizes larger than 30. Now, the exponential growth of the number of computational basis states suppresses the performance of sampling. For problem size 31, the VQE achieves a better approximation ratio for roughly 95\% of instances, corresponding to an average difference in approximation ratio of around 0.05. After around 100 iterations and for problem sizes larger than 40, the VQE always performs better than sampling on average.

\begin{figure}
    \begin{subfigure}{0.48\textwidth}
        \centering
        \includegraphics[width=\textwidth]{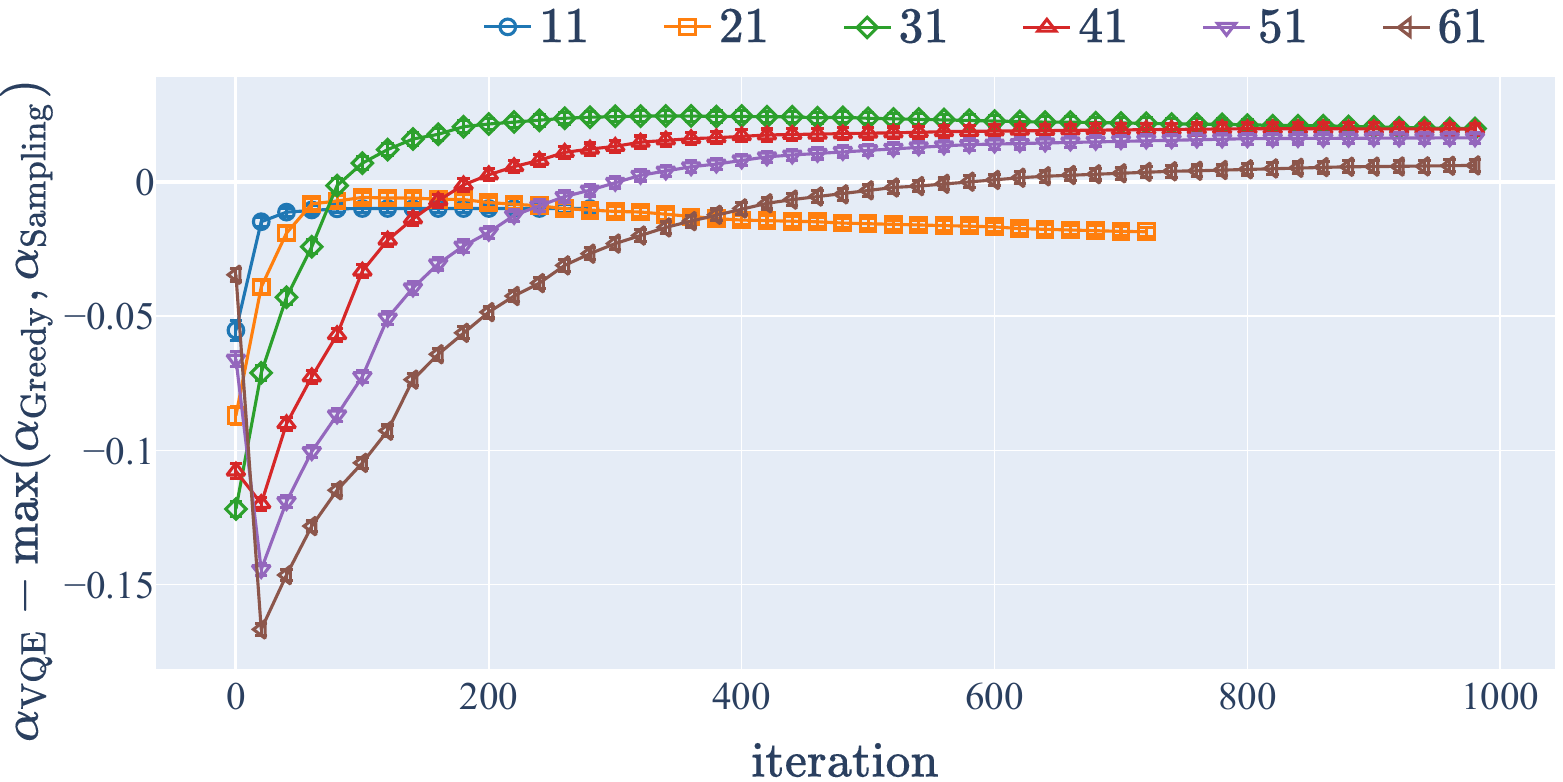}
        \caption{Mean and SEM of the difference in approximation ratio. The greedy algorithm converges to local minima very quickly. Then the VQE catches up and eventually converges to local minima closer to the optimal value.}
        \label{fig:approx_ratio_greedy}
    \end{subfigure}
    \hfill
    \begin{subfigure}{0.48\textwidth}
        \centering
        \includegraphics[width=\textwidth]{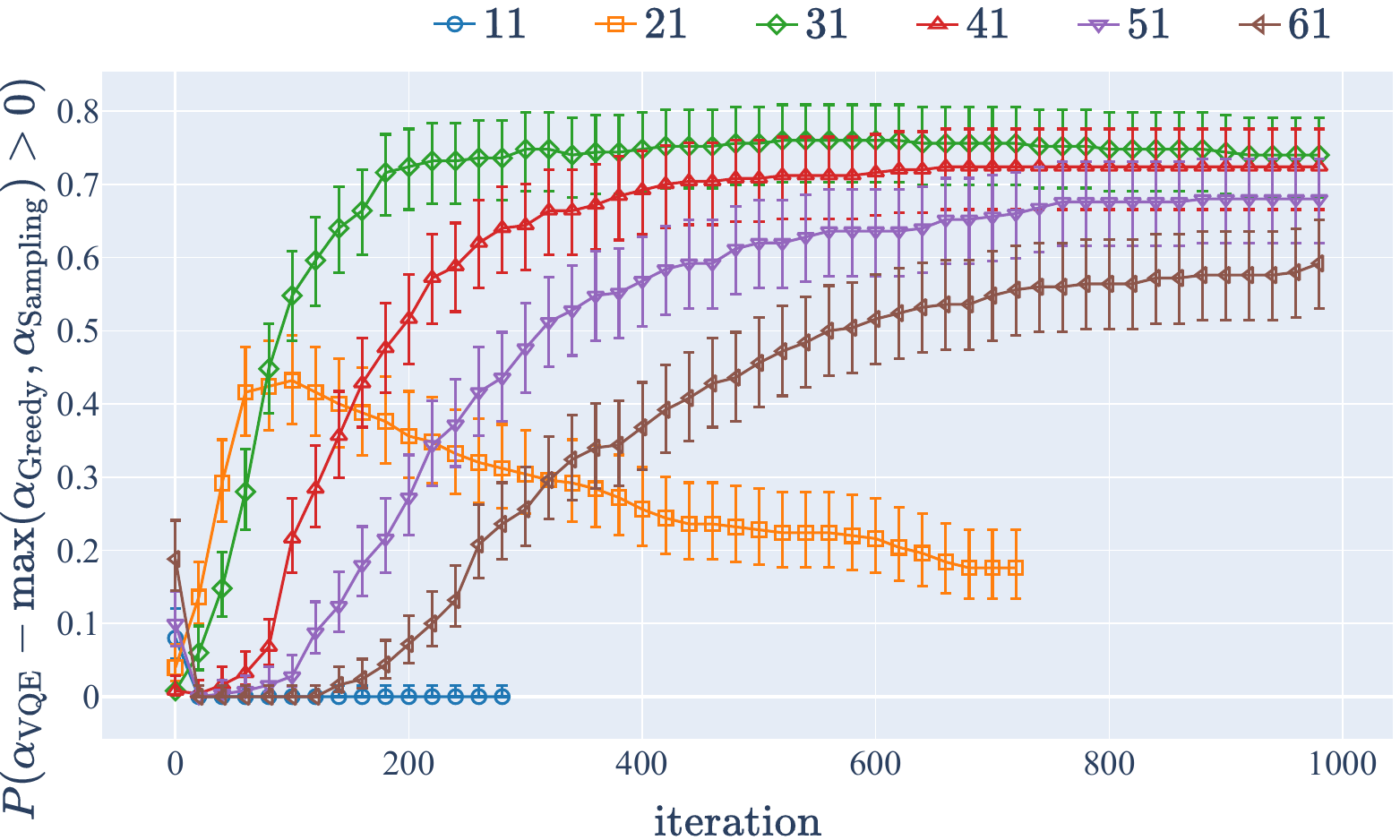}
        \caption{Mean probability and 95\% confidence interval of the VQE achieving a better approximation ratio than sampling and the greedy algorithm.}
        \label{fig:better_approx_ratio_greedy}
    \end{subfigure}
    \hfill
    \caption{Average difference in performance metrics of the VQE and the best of both the sampling and the greedy algorithm as function of the number of iterations for different problem sizes indicated by the different markers. As a guide for the eye, the different data points for each problem size are connected with a line. 
    } 
    \label{fig:vqe_minus_greedy}
\end{figure}

Sampling is the simplest way to generate solution candidates. In a next step, we compare the VQE to a simple set of rules to generate computational basis states that define the greedy algorithm described in Sec.~\ref{sec:greedy}. To avoid misleading conclusions, we compare the VQE with the best performing of both the greedy algorithm and the sampling approach. In particular, we only observe that the sampling algorithm is better than the greedy approach for problem size 11 and up to some iterations for problem size 21. The approximation ratios achieved by the VQE minus the best one obtained by the greedy algorithm or by sampling are shown in Fig.~\ref{fig:vqe_minus_greedy}.
For problem sizes 11 we see again the regime where the VQE does not perform better than sampling. Up to the resolution given by the number of measurements per iteration that we use to compare the algorithms, the greedy algorithm does not perform better than sampling neither. This is because the space of $2^{11}$ solution candidates is very small, which makes sampling a suitable strategy to solve the problem. For this problem size, it does not make sense to compare the performance of the VQE and the greedy algorithm, because both perform worse than sampling. For problem size 21, the small advantage of the VQE over sampling for a few early iterations is taken by the greedy algorithm (see Fig.~\ref{fig:vqe_minus_greedy}(a)). This means that up to problem size 21 the VQE performs strictly worse than the two classical algorithms.

For problem sizes larger than 30, we observe a characteristic behavior of the algorithms. The greedy algorithm converges to local minima very quickly, leading to a better approximation ratio than the early iterations of the VQE. Then, the VQE slowly catches up and eventually converges to better local minima than the greedy algorithm on average. The iteration count for which the VQE matches the performance of the greedy algorithm increases with the problem size from around 100 iterations for size 31 to around 600 for size 61. The maximal advantage of the VQE over the greedy algorithm, on the other hand, decreases with the problem size, as Fig.~\ref{fig:vqe_minus_greedy} reveals. For size 31, the probability of the VQE for achieving a higher approximation ratio than the greedy algorithm reaches up to around 75\% compared to around 60\% for size 61. This corresponds to an average difference in approximation ratio of around 0.025 for size 31 and of around 0.005 for size 61 (see Fig.~\ref{fig:vqe_minus_greedy}(a)). 

\begin{figure}
    \begin{subfigure}{0.48\textwidth}
        \centering
        \includegraphics[width=\textwidth]{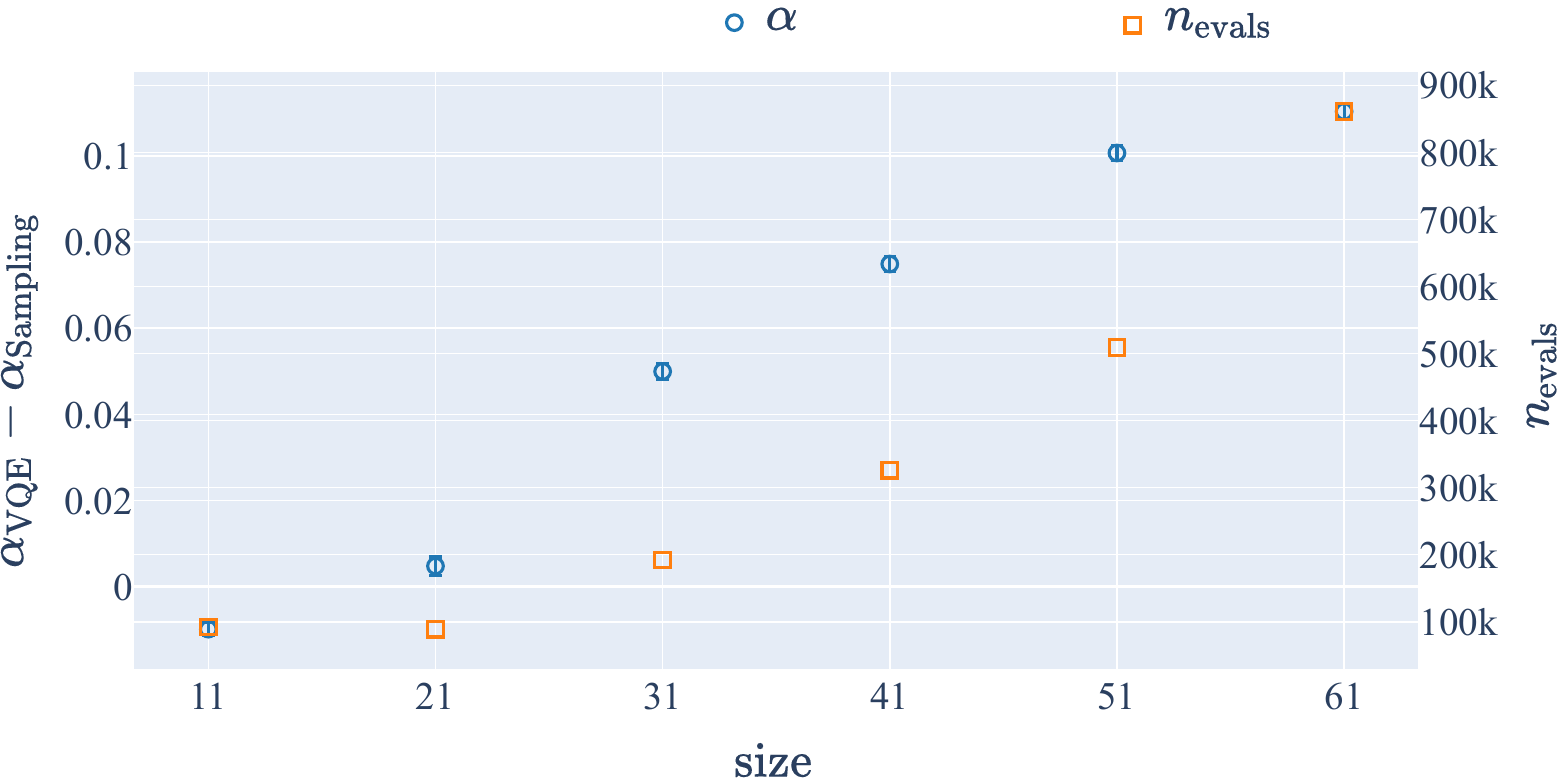}
        \caption{Difference between the VQE and sampling. The advantage of the VQE over sampling increases with problem size.}
        \label{fig:max_diff_sampling}
    \end{subfigure}
    \hfill
    \begin{subfigure}{0.48\textwidth}
        \centering
        \includegraphics[width=\textwidth]{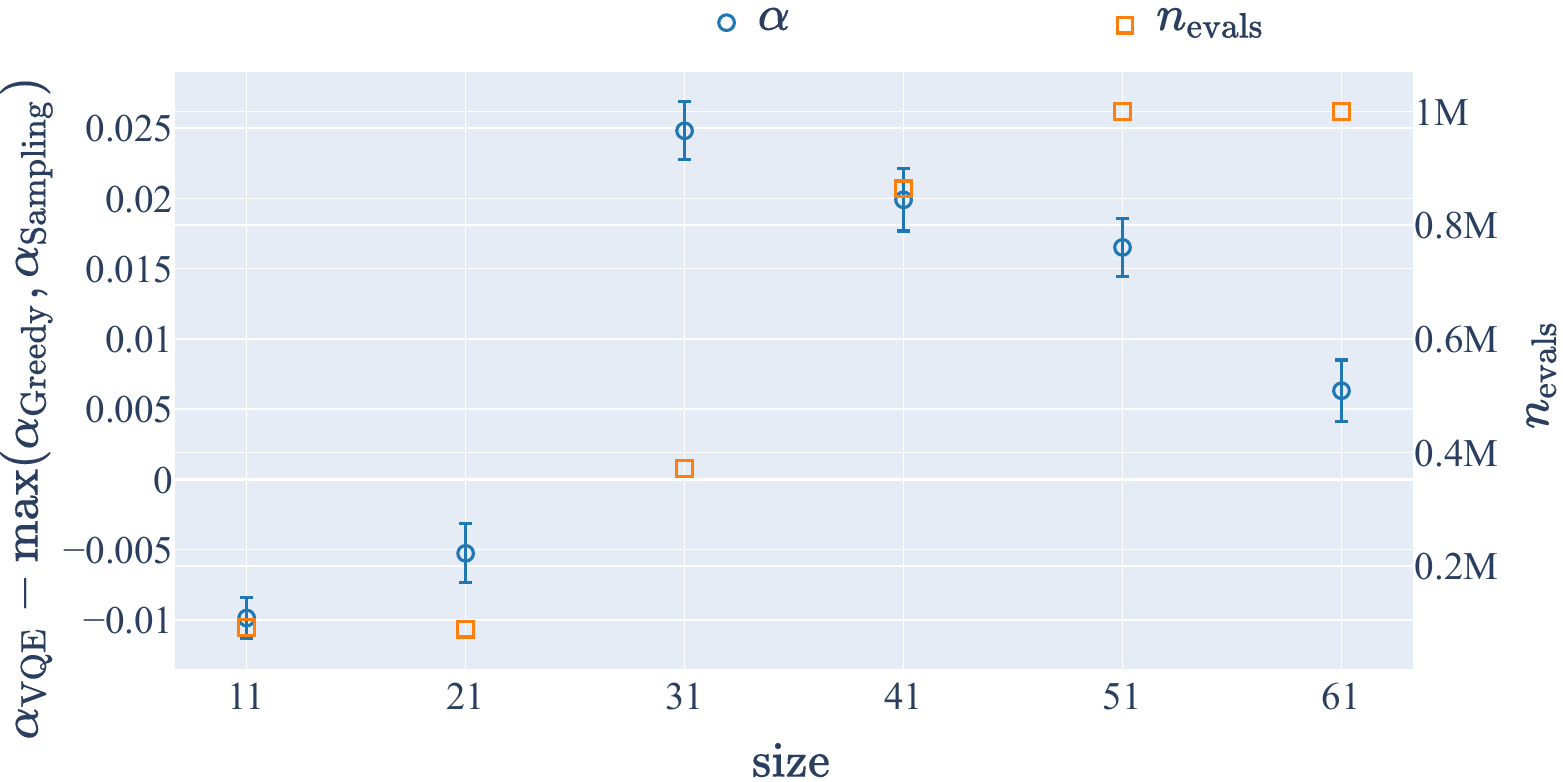}
        \caption{Difference between the VQE and the greedy algorithm. For size 11 and 21 VQE performs worse than the other algorithms. Then, the VQE shows an advantage over the greedy algorithm that decreases with problem size.}
        \label{fig:max_diff_greedy}
    \end{subfigure}
    \hfill
    \caption{Average difference in approximation ratio (blue dots and left $y$-axis) and the corresponding number of objective function evaluations (orange squares and right $y$-axis) for the iteration of the VQE where the distance is maximal. Like in the previous figure, we compare the VQE to sampling in regions where the greedy algorithm does not perform better than sampling. This is only the case for size 11.}
    \label{fig:max_diff}
\end{figure}

In Fig.~\ref{fig:max_diff}, we show the average difference between the VQE and sampling (panel (a)) and the VQE and the better one of sampling and the greedy algorithm (panel (b)) for the iteration of the VQE, where the average difference is maximal. Furthermore, we determine the number of evaluations of the objective function $n_{\mathrm{evals}}$ to achieve the maximal difference. We recall that we set the maximal number of objective function evaluations to $N_{\mathrm{evals}}=10^6$ for our numerical experiments. 
Focusing on the case of sampling in Fig.~\ref{fig:max_diff}(a) first, we see that for 11 variables, the smallest problem size considered in our study, the VQE performs worse than sampling on average. Moving to the next larger problem of size 21, the VQE shows a small advantage over sampling. For the problem sizes considered in our study, we observe a monotonous increase of the advantage of the VQE over sampling. The greatest advantage over sampling is achieved for size 61 with a average difference in approximation ratio of around 0.11. The corresponding number of objective function evaluations also increases monotonically from around $10^5$ for size 21 to around $8.5\times10^5$ for size 61.  

Comparing the VQE to the better one of sampling and the greedy algorithm, Fig.~\ref{fig:max_diff}(b) reveals the opposite behavior, i.e., a decreasing advantage of the VQE with problem size. For size 21 the VQE shows an average approximation ratio that is around 0.005 smaller than the one achieved by the greedy algorithm. Then, the advantage of the VQE over the greedy algorithm peaks for size 31 at around 0.025 and reduces to around 0.005 for size 61. 
Like in the case of sampling, the corresponding number of objective function evaluations increases monotonically. For the two largest problems considered in our study with sizes 51 and 61 this number reaches our limit of $10^6$. This means that additional iterations of the VQE could further increase the advantage. However, Fig.~\ref{fig:vqe_minus_greedy} suggests that the advantage has converged closely to its optimal value. 

\begin{figure}
    \begin{subfigure}{0.48\textwidth}
        \centering
        \includegraphics[width=\textwidth]{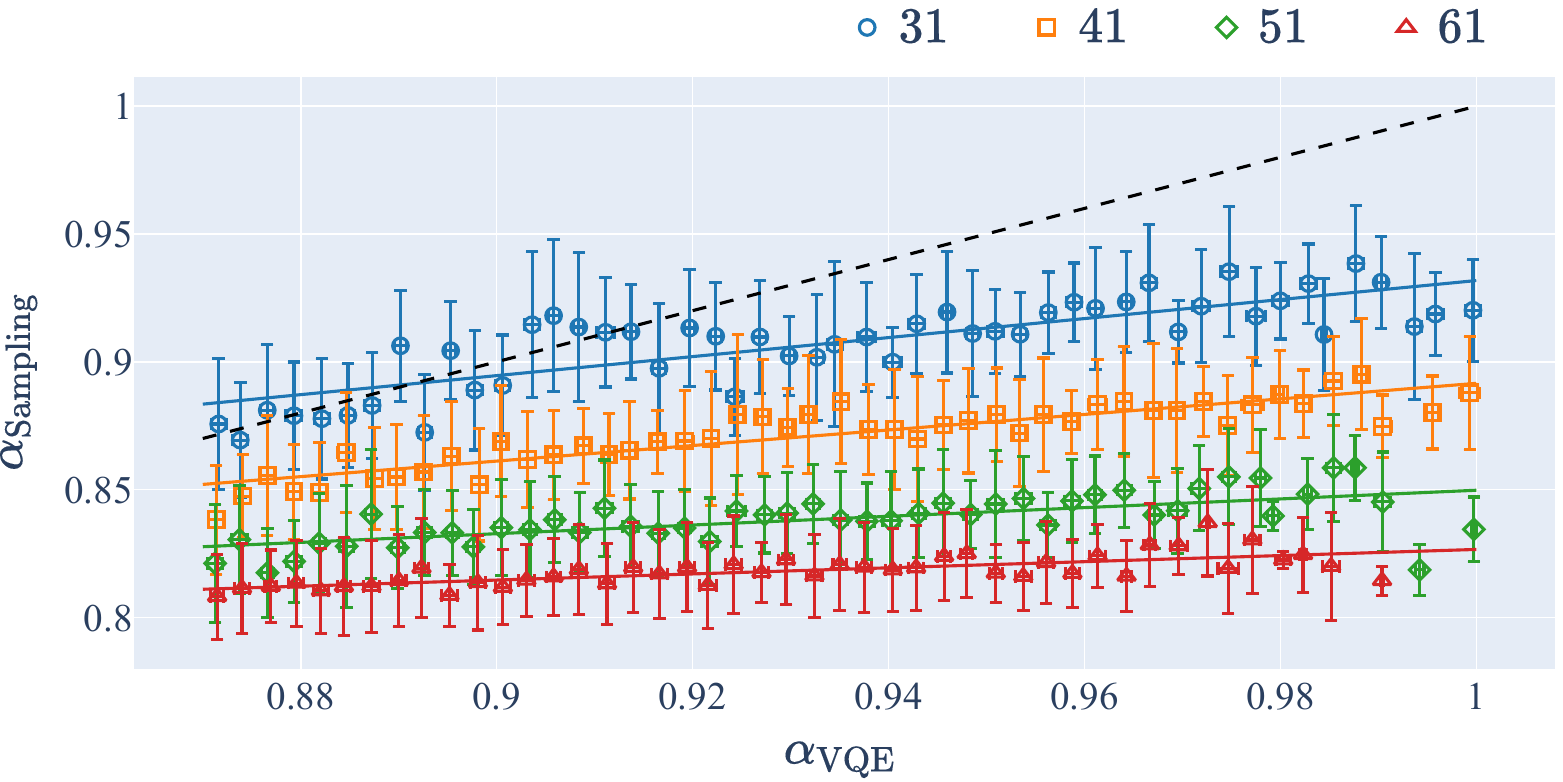}
        \caption{Correlations between the VQE and sampling. The performance of sampling depends strongly on the problem size.}
        \label{fig:corr_sampling}
    \end{subfigure}
    \hfill
    \begin{subfigure}{0.48\textwidth}
        \centering
        \includegraphics[width=\textwidth]{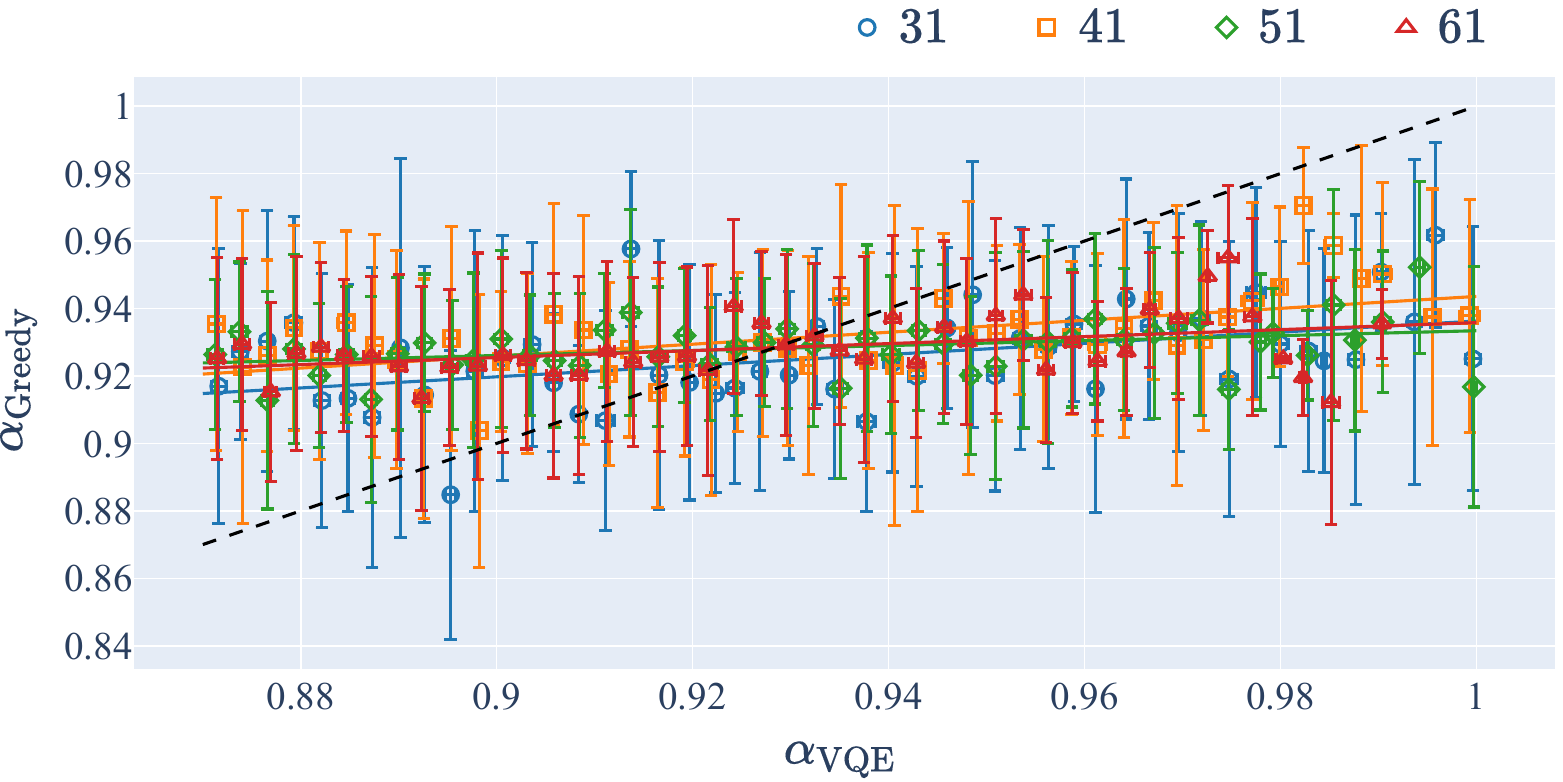}
        \caption{No correlation between the VQE and the greedy algorithm. The average performance of the greedy algorithm is constant for the problem sizes considered in this plot.}
        \label{fig:corr_greedy}
    \end{subfigure}
    \hfill
    \caption{Mean and standard deviation of the approximation ratio of sampling and the greedy algorithm in dependence on the approximation ratio of the VQE using a binned statistic.}
    \label{fig:corr}
\end{figure}

As a final step of our analysis, we extend the commonly studied average case by investigating the correlation of the performance of the algorithms by instance. In Fig.~\ref{fig:corr}, we plot the approximation ratio of sampling and of the greedy algorithm as a function of the approximation ratio of the VQE. We use binned statistics as explained in Sec.~\ref{sec:metrics} and analyze the regime where the VQE performs better than the worst-case guarantee of the efficient classical GW algorithm. Furthermore, we exclude the problem sizes 11 and 21 where the VQE performs strictly worse than sampling or the greedy algorithm. 

The approximation ratio of sampling and the VQE show a mild correlation that depends on the problem size.
This correlation is expected because sampling is a key component of the VQE algorithm. Since sampling does not depend on initial points the varying performance of sampling is solely a consequence of the varying hardness with respect to sampling of the randomly generated Max-Cut instances. Intuitively, sampling performs better for instances with many solution candidates close to the optimal solution in the objective function value.
The performance of sampling drops quickly with the problem size for the full range of approximation ratios of the VQE and the correlation of the performance decreases with the problem size. This suggests that the variance of the hardness of the instances with respect to sampling declines with problem size. The varying performance of the VQE is then explained by its initial points. 

A completely different picture arises for the correlation of the performance of the greedy algorithm and the VQE. The standard deviation of the approximation ratio of the greedy algorithm for each bin is comparable to the full range of the approximation ratios we study. This means that there is no correlation between the two algorithms, highlighting the fact that the VQE and the greedy algorithm rely on fundamentally different mechanisms. The linear fit of the approximation ratio shows that the average approximation ratio achieved by the greedy algorithm is independent of the problem size for the sizes shown in Fig.~\ref{fig:corr}.

\section{\label{sec:dicussion}Discussion and outlook}
In this study we have conducted extensive numerical experiments to demonstrate intuitive benchmarks for VQAs beyond proof of principles. For small instances of the frequently used Max-Cut problem on 3-regular graphs, we observed that randomly sampling solutions yields high approximation ratios. Thus, it is crucial to perform resource-aware comparisons between VQAs and sampling in this regime. 
For the specific VQE of our study, we found that it performs better than sampling only for problems with more than around 30 variables. This implies that it is necessary to probe larger problems to make meaningful statements about the algorithm. Advanced simulation methods, such as tensor networks, are able to simulate shallow quantum circuits for many qubits with reasonable computational effort. 

Another intuitive benchmark is the comparison of VQAs to fixed set of rules to generate samples to evaluate the objective function. In our study, these rules define a greedy algorithm that yields a constant average approximation ratio for the larger problems. This behavior bounds the regime where the VQA could have an advantage from above, complementing the bound from below given by the performance of sampling. 
By starting the greedy algorithm from the same initial point as the VQE in the sense defined in Sec.~\ref{sec:greedy} we were able to demonstrate that good initial points for the VQE are not necessarily good initial points for the greedy algorithm. For an approach for finding good initial points for VQAs, we refer to~\cite{Chai2024}. 

Our benchmarks rely on the property that the solutions of combinatorial problems can be encoded in computational basis states. Since VQAs are mostly applied to physical problems where this is not the case, it would be insightful to extend our benchmarks to these kinds of problem. For example, many VQAs for physical systems aim to measure the expectation values of observables. A common classical tool for such computations would be Markov Chain Monte Carlo (MCMC) methods. It would therefore be highly interesting to see how our observations that minimum complexity is required for VQAs to be have a potential advantage correlates with metrics that describe the hardness of MCMC methods, such as autocorrelation. Similarly, a VQA might be used to approximate the state of a physical system and once this state is found, observables can be measured. Hence, the main computational effort for the VQA leads into the state resolution procedure which is independent of the observable to be measured while many MCMC calculations are observable dependent. This raises the question whether there are classes of observables with expectation values that are easier to compute using VQA over MCMC and vice versa. 

Ultimately, the economic efficiency of an algorithm -- defined by the quality of its solution and the time and financial resources it requires -- decides whether an algorithm is useful in practice. We once again want to highlight that the instances typically used to benchmark quantum algorithms are trivial for commercial classical solvers. However, one has to keep in mind that VQAs are at an early stage of quantum computation that is rapidly evolving in both algorithmic design and hardware development. We hope that our work provides a step towards meaningful benchmarks of quantum algorithms beyond proof-of-principles.

\begin{acknowledgments}
This work is supported by: 
i) the Einstein Research Unit -``Perspectives of a quantum digital transformation: Near-term quantum computational devices and quantum processors'', 
ii) the European Union’s Horizon Europe research and innovation funding programme under the ERA Chair scheme with grant agreement No. 101087126 and under the European Research Council (ERC) with grant agreement No. 787331,
iii) the Helmholtz Association -``Innopool Project Variational Quantum Computer Simulations (VQCS)'',
and iv) the Ministry of Science, Research and Culture of the State of Brandenburg within the Centre for Quantum Technologies and Applications (CQTA). \raisebox{-1.0em}{\includegraphics[width = 0.05\textwidth]{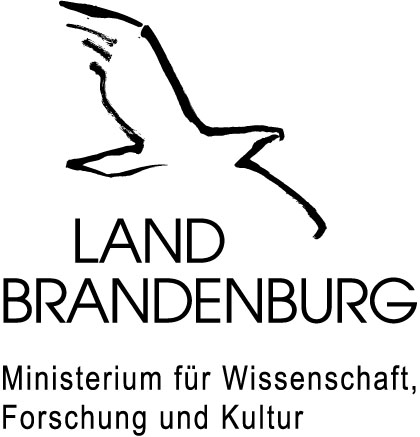}}
\end{acknowledgments}

\bibliography{apssamp}

\end{document}